\let\accentvec\vec
\documentclass{llncs}
\usepackage{llncsdoc}

\let\vec\accentvec

\usepackage{fancyvrb}
\usepackage{graphicx}
\usepackage{amsmath}
\usepackage{color}
\usepackage{algorithm}
\usepackage{tikz}
\usetikzlibrary{shapes,shapes.multipart,arrows}
\usepackage{algorithmic}
\usepackage{amsfonts}
\usepackage{multirow}
\usepackage[bookmarks=false]{hyperref}
\usepackage{paralist}
\usepackage{comment}
\usepackage{float}

\definecolor{xorange}{cmyk}{0,0.65,1,0.09}
\definecolor{xgreen}{rgb}{0.66, 0.77, 0.5}
\definecolor{xblue}{rgb}{0.07, 0.37, .40}
\definecolor{xred}{rgb}{.9 .1 .3}

\newcommand{\upone}{\vdash_1}
\newcommand{\emptys}{\bot}

\newcommand{\rat}{\textsc{RAT}}

\newcommand{\CaC}{C\&C}

\title{Solving and Verifying the boolean Pythagorean Triples problem via Cube-and-Conquer}
\author{Marijn J. H. Heule, Oliver Kullmann, and Victor W. Marek}
\institute{The University of Texas at Austin, Swansea University, and University of Kentucky}

\newcommand{\NN}{\mathbb{N}}
\DeclareMathOperator{\ptn}{Ptn}
\DeclareMathOperator{\var}{var}

\iffalse
\newcommand{\mfalse}{\textup{\textbf{f}}}
\newcommand{\mtrue}{\textup{\textbf{t}}}
\else
\newcommand{\mfalse}{\ensuremath 0}
\newcommand{\mtrue}{\ensuremath 1}
\fi
\newcommand{\ta}{\tau}

\begin{document}

\maketitle

\begin{abstract}
The \emph{boolean Pythagorean Triples problem} has been a longstanding
open problem in Ramsey Theory: Can the set
$\NN = \{1,2,\dots\}$ of natural numbers be divided into two parts,
such that no part contains a triple $(a,b,c)$ with $a^2 + b^2 = c^2$ ?
A prize for the solution was offered by Ronald Graham over two decades ago.
We solve this problem, proving in fact the impossibility, by using the
\emph{Cube-and-Conquer} paradigm, a hybrid SAT method for
hard problems, employing both look-ahead and CDCL solvers.
An important role is played by dedicated look-ahead
heuristics, which indeed allowed to solve the problem on a cluster with 800 cores in about 2 days.
Due to the general interest in this mathematical problem, our result
requires a formal proof. Exploiting recent progress
in unsatisfiability proofs of SAT solvers, we produced and verified
a proof in the DRAT format, which is almost 200 terabytes in size.
From this we extracted and made available a compressed certificate of 68 gigabytes, that
allows anyone to reconstruct the DRAT proof for checking.
\end{abstract}

\section{Introduction}

Propositional satisfiability (SAT, for short) is a formalism that allows for
representation of all finite-domain constraint satisfaction problems.
Consequently, all decision problems in the class NP, as well as all search
problems in the class FNP~\cite{Co71,Karp,Levin,Garey-Jo}, can be polynomially reduced to SAT.
Due to great progress with {\em SAT solvers}, many practically important problems are solved using such reductions. SAT is especially an important tool in
hardware verification, for example model checking~\cite{Clarke99} and reactive systems checking.
In this paper we are, however, dealing with a different application of SAT,
namely as a tool in computations of configurations in a part of Mathematics
called {\em Extremal Combinatorics}, especially \emph{Ramsey theory}. In this area, the researcher attempts to
find various configurations that satisfy some combinatorial conditions, as well
as values of various parameters associated with such configurations~\cite{Zha09HBSAT}. 

One important result of Ramsey theory, the van der Waerden
Theorem~\cite{vdW27}, has been studied by the SAT community, started by~\cite{DMT04}.
That theorem says that for all natural
numbers $k$ and $l$ there is a number $n$, so that whenever the integers
$1,\dots,n$ are partitioned into $k$ sets, there is a set containing an arithmetic
progression of length $l$. 
A good deal of
effort has been spent on specific values of the corresponding number theoretic
function, $\mbox{vdW}(k,l)$. Two results on specific values:
$\mbox{vdW}(2,6) =1132$ and $\mbox{vdW}(3,4)=293$, were obtained by M.~Kouril
\cite{KP08,Ko12} using specialized FPGA-based
SAT solvers.  
Other examples 
include the Schur Theorem \cite{Sch16} on
sum-free subsets, its generalization known as Rado's Theorem \cite{Rado70},
and a generalization of van der
Waerden numbers~\cite{AhmedKullmannSnevily2011VdW3kArt}. 
In this paper we investigate two areas:
\begin{enumerate}
\item  We show the
``boolean Pythagorean triples partition theorem'' (Theorem \ref{thm:bPyth}), or colouring of Pythagorean triples, an analogue of Schur's Theorem.
\item We develop methods to compute numbers in Ramsey theory by SAT solvers.
\end{enumerate}

A triple  $(a,b,c) \in \NN^3$ is called {\em Pytha\-go\-rean} if $a^2 + b^2 = c^2$.
If for some $n > 2$ all partitions of the set $\{1,\dots,n\}$ into two parts contain a
Pythagorean triple in at least one part, then that property holds for all such partitions of
$\{1,\dots,m\}$ for $m \ge n.$ A partition by Cooper and Overstreet~\cite{CO15} of the set $\{1,\dots,7664\}$ into two parts, with no part containing
a Pythagorean triple, was previously the best result, thereby improving on earlier lower bounds~\cite{Cooper1344,Kay1514,Myers2015}.


\begin{theorem}\label{thm:bPyth}
  The set $\{1,\dots,7824\}$ can be partitioned into two parts, such that no part contains a Pythagorean 
  triple
, while this is impossible for $\{1,\dots,7825\}$.
\end{theorem}

Graham repeatedly offered a prize of \$100 for proving such a theorem, and the problem is explicitly stated in [10].
To emphasize, the situation of Theorem \ref{thm:bPyth} is not as in previous applications of SAT to Ramsey theory, 
where SAT only ``filled out the numerical details'', but the existence of these numbers was not known 
(and as such is a good success of Automated Theorem Proving). 
It is natural to generalize our problem in a manner similar to the Schur Theorem:

\begin{conjecture}\label{con:Pyth}
For every $k \ge 1$ there exist $\ptn(k) \in \NN$ (the ``Pythagorean triple
number''), such that $\{1,\dots,\ptn(k)-1\}$ can be partitioned
into $k$ parts with no part containing a Pythagorean triple, while this is
impossible for $\{1,\dots,\ptn(k)\}$.
\end{conjecture}

We prove Theorem \ref{thm:bPyth} by considering two SAT problems.
One showing that $\{1,\dots,7824\}$ can be partitioned into two parts such 
that no part contains a Pythagorean triple (i.e., the case $n=7824$ is satisfiable).
The other one showing that any partitioning of $\{1,\dots,7825\}$ into two parts 
contains a Pythagorean triple (i.e., the case $n=7825$ is unsatisfiable). 
Now a Pythagorean triple-free partition for $n = 7824$ is checkable in a second, 
but the {\em absence} of such a partition for $n=7825$ requires a more ``durable proof'' than
just the statement that we run a SAT solver (in some non-trivial fashion!) which
answered UNSAT --- to become a mathematically accepted theorem, our assertion for
$n = 7825$ carries a stronger burden of proof. Fortunately, the SAT community
has spent a significant effort to develop techniques that allow
to extract, out of a failed attempt to get a satisfying assignment,
an actual {\em proof of the unsatisfiability}.

It is worth noting the similarities and differences to the endeavours of extending mathematical arguments into actual {\em formal proofs}, using tools like {\em Mizar}
\cite{Mizar} and {\em Coq} \cite{Coq}. Cases, where intuitions (or convictions) about completeness of mathematical arguments fail, are known~\cite{Voy14}. So T.~Hales in his project {\em
flyspeck} \cite{flyspeck} extracted and
verified his own proof of the {\em Kepler Conjecture}. Now the core of the argument in such examples has been constructed by mathematicians. Very different from that, the proofs
for unsatisfiability coming from SAT solvers are, from a human point of view, a giant heap of random information (no direct understanding is involved). 
But we don't need to search for the proof --- the present generation of SAT solvers supports emission of unsatisfiability proofs and 
standards for such proofs exist \cite{Wetzler14}, 
as well as checkers that the proof is valid. However the proof that we will encounter
in our specific problem is of very large size. In fact, even {\em storing} it
is a significant task, requiring significant compression. We will tackle these problems in this paper.

\section{Preliminaries}

\paragraph{\bf CNF Satisfiability.}
For a Boolean variable $x$,
there are two \emph{literals}, the positive literal $x$ 
and the negative literal $\bar x$.
A \emph{clause}  is a finite set of literals; so it may contain complementary
literals, in which case the clause is tautological. The empty clause is denoted by $\emptys$.
If convenient, we write
a clause as a disjunction of literals. Since a clause is a set, no literal
occurs several times, and the order of literals in it does not matter.
A (CNF) \emph{formula} is a conjunction of clauses, and thus clauses may occur several
times, and the order of clauses does matter; in many situations these
distinctions can be ignored, for example in semantical situations, and then
we consider in fact finite sets of clauses.

A \emph{partial assignment}
is a  function $\ta$ that maps a finite set of literals
to $\{\mfalse,\mtrue\}$, such that for $v \in \{\mfalse,\mtrue\}$ holds
$\ta(x) = v$ if and only if $\ta(\bar{x}) = \lnot v$.
A clause $C$ is satisfied by $\ta$ if  $\ta(l) = \mtrue$ for some
literal $l\in C$, while $\ta$ satisfies a formula $F$ if it satisfies every 
clause in $F$.
If a formula $F$ contains $\emptys$, then $F$ is unsatisfiable.
A formula $F$ \emph{logically implies} another formula $F'$, denoted by 
$F \models F'$, if every satisfying assignment for $F$ also satisfies $F'$. 
A transition $F \leadsto F'$ is \emph{sat-preserving}, if
either $F$ is unsatisfiable or both $F, F'$ are satisfiable, while the
transition if \emph{unsat-preserving} if either $F$ is satisfiable or both $F, F'$ 
are unsatisfiable. Stronger, $F, F'$ are \emph{satisfiability-equivalent}
if both formulas are satisfiable or both unsatisfiable, that is, iff the transition 
$F \leadsto F'$ is both sat- and unsat-preserving. We note that if 
$F \models F'$, then $F \leadsto F'$ is sat-preserving, and that $F \leadsto F'$
is sat-preserving iff $F' \leadsto F$ is unsat-preserving. Clause addition
is always unsat-preserving, clause elimination is always sat-preserving.

\paragraph{\bf Resolution and Extended Resolution.}

The resolution rule (see \cite[Subsections 1.15-1.16]{FM09HBSAT}) infers from two clauses $C_1 = (x \lor
a_1\lor \ldots \lor a_n)$ and $C_2 = (\bar x \lor b_1 \lor \ldots \lor
b_m)$ the \emph{resolvent} $C = (a_1\lor \ldots \lor a_n \lor
b_1 \lor \ldots \lor b_m)$, by resolving on variable
$x$.
$C$ is logically implied by any formula
containing $C_1$ and $C_2$.
For a given CNF formula $F$, the \emph{extension
rule}~\cite{Tseitin:complexity} allows one to iteratively add
definitions of the form $x := a \land b$ by adding the {\em extended resolution clauses}
$(x \lor \bar a \lor \bar b) \land (\bar x \lor a) \land (\bar x \lor
b)$ to $F$, where $x$ is a new variable and $a$ and $b$ are literals
in the current formula. The addition of these clauses is sat-equivalent.
\vspace{-4pt}

\paragraph{\bf Unit Propagation.}

For a CNF formula $F$, \emph{unit propagation} simplifies $F$ based 
on unit clauses;
that is, it repeats the following until fixpoint:
if there is a unit clause $\{l\} \in F$, remove all clauses that
contain the literal $l$ from the set $F$ and
remove the literal $\bar l$ from the remaining clauses in $F$.
This process is sat-equivalent.
%
If unit propagation on formula $F$ produces complementary 
units $\{l\}$ and $\{\bar l\}$, we say that unit propagation
\emph{derives a conflict} and write $F\upone \emptys$
(this relation also holds
if $\emptys$ is already in $F$).
%

Ordinary resolution proofs (or ``refutations'' -- derivations of the empty clause) just add resolvents. This is too inefficient, and is extended via unit propagation as follows. For a clause $C$ let $\neg C$ denote the conjunction of unit clauses that falsify all literals in $C$.
A clause $C$ is an \emph{asymmetric tautology} with respect to a CNF formula $F$ if
$F \land \neg C \upone \emptys$. This is equivalent to
the clause $C$ being derivable from $F$ via {\em input resolution}~\cite{He74}:
a sequence of resolution steps using for every resolution step at least one
clause of $F$. So addition of resolvents is generalised by addition of
asymmetric tautologies (where addition steps always refer to the
current (enlarged) formula, the original axioms plus the added clauses).
Asymmetric tautologies,
also known as {\em reverse unit propagation} (RUP)
clauses, are the most common learned 
clauses in {\em conflict-driven 
clause learning} (CDCL) SAT solvers
(see \cite[Subsection 4.4]{MSLM09HBSAT}).
This extension is irrelevant from the proof-complexity point of view, but for practical applications exploitation of the power of fast unit propagation algorithms is essential.

\paragraph{\bf RAT clauses.}

We are seeking to add sat-preserving clauses beyond logically implied clauses.
The basic idea is as follows (proof left as instructive exercise):
\begin{lemma}\label{lem:genrat}
  Consider a formula $F$, a clause $C$ and a literal $x \in C$. 
  If for all $D \in F$ such that $\bar{x} \in F$ it holds that $F \models C \cup (D \setminus \{\bar x\})$, then addition of $C$ to $F$ is sat-preserving.
\end{lemma}

In order to render the condition $F \models C \cup (D \setminus \{\bar x\})$ polytime-decidable, 
we stipulate that the right-hand clause must be derivable by input resolution:
\begin{definition}[\cite{rules}]\label{def:rat}
  Consider a formula $F$, a clause $C$ and a literal $x \in C$ (the ``pivot''). We say that $C$ has $\rat$ (``Resolution asymmetric tautology'') on $x$ w.r.t.\ $F$ if for all $D \in F$ with $\bar x \in D$ holds that
$
F \land \neg (C \cup (D \setminus \{\bar x\})) \upone \emptys
$.
\end{definition}
By Lemma \ref{lem:genrat}, addition of RAT-clauses is sat-preserving.
Every non-empty asymmetric tautology $C$ for $F$ has RAT on any $x \in C$ w.r.t.\ $F$. 
It is also easy to see that the three extended resolution clauses are RAT clauses 
(using the new variable for the pivot literals). All preprocessing, inprocessing, and solving techniques in
state-of-the-art SAT solvers can be expressed in terms of addition and
removal of $\rat$ clauses~\cite{rules}.

\vspace{-4pt}

\section{Proofs of Unsatisfiability}

A {\em proof of
unsatisfiability} (also called a {\em refutation}) for a formula $F$
is a sequence of sat-preserving transitions which ends with some formula
containing the empty clause.
There are currently two prevalent types of unsatisfiability proofs:
\emph{resolution proofs} and {\em clausal proofs}. Both do not display the
sequence of transformed formulas, but only list the axioms (from $F$) and
the additions and (possibly) deletions.
Several formats have
been designed for resolution
proofs~\cite{Zhang:2003,Een:2003,picosat} (which only add clauses), but they all share the same
disadvantages. Resolution proofs are often huge, and it is hard to
express important techniques, such as conflict clause minimization, with
resolution steps. Other techniques, such as bounded variable addition~\cite{reencoding},
cannot be polynomially-simulated by resolution at all. Clausal proof
formats~\cite{Wetzler14,Gelder:2008,Heule:2013:RAT} are syntactically
similar; they involve a sequence of clauses that are claimed to be
sat-preserving, starting with the given formula. But now we might add
clauses which are not logically implied, and we also might remove clauses
(this is needed now in order to enable certain additions, which might depend
on global conditions).

\begin{definition}[\cite{Wetzler14}]\label{def:DRAT}
  A \emph{DRAT proof} (``Deletion Resolution Asymmetric Tautology'') for a
  formula $F$ is a sequence of additions and deletions of clauses, starting
  with $F$, such that each addition is the addition of a RAT clause
  w.r.t.\ the current formula (the result of additions and deletions up
  to this point), or, in case of adding the empty clause, unit-clause
  propagation on the current formula yields a contradiction.
  A \emph{DRAT refutation} is a DRAT proof containing $\emptys$.
\end{definition}

DRAT refutations are 
correct proofs of unsatisfiability
(based on Lemma \ref{lem:genrat} and the fact, that deletion of clauses is
always sat-preserving; note that Definition \ref{def:DRAT} allows unrestricted deletions). Furthermore they are checkable in cubic time. Since
the proof of Lemma \ref{lem:genrat} is basically the same as the proof for
\cite[Lemma 4.1]{Ku96c}, by adding unit propagation appropriately one can
transfer \cite[Corollary 7.2]{Ku96c} and prove that the power of DRAT
refutations is up to polytime transformations the same as the power of
Extended Resolution. 

\begin{figure}[t]
\centering
~
\begin{minipage}[t]{.18\textwidth}
\centering
\vspace{-67pt}
{CNF formula}
\begin{Verbatim}[frame=single]
 p cnf 4 8
  1  2 -3 0
 -1 -2  3 0
  2  3 -4 0
 -2 -3  4 0
 -1 -3 -4 0
  1  3  4 0
 -1  2  4 0
  1 -2 -4 0
\end{Verbatim}
\end{minipage}
~
\begin{minipage}[t]{.18\textwidth}
\centering
\vspace{-67pt}
{DRAT proof}
\vspace{-0.048cm}
\begin{Verbatim}[frame=single]

   -1 0
 d -1 2 4 0
    2 0
    0




\end{Verbatim}
\end{minipage}
\hfill
\begin{minipage}{.57\textwidth}
\vspace{-20pt}
\caption{Left, a formula in DIMACS CNF format, the conventional
  input for SAT solvers which starts with {\tt p cnf} to denote
  the format, followed by the number of variables and the number of
  clauses.  Right, a DRAT refutation for that formula.
  Each line in the proof is either an addition step (no prefix) or a
  deletion step identified by the prefix ``{\tt d}''. Spacing
  is used to improve readability.  Each clause in the proof
  must be a RAT clause using the first literal as pivot, or the
  empty clause as an asymmetric tautology.}
\label{figure:input-formats}
\end{minipage}
\end{figure}

\begin{example}
 Figure~\ref{figure:input-formats} shows an example
DRAT refutation.
Consider the CNF formula $F=(a \lor b \lor \bar c) \land (\bar
a \lor \bar b \lor c) \land (b \lor c \lor \bar d) \land (\bar
b \lor \bar c \lor d)\land (a \lor c \lor d) \land (\bar a \lor \bar
c \lor \bar d) \land (\bar a \lor b \lor d) \land ( a \lor \bar
b \lor \bar d)$, shown in DIMACS format in
Fig.~\ref{figure:input-formats} (left), where 1 represents $a$, 2 is $b$,
3 is $c$, 4 is $d$, and negative numbers represent negation.  
The first clause in the proof, $(\bar a)$, is a RAT clause with respect to 
$F$ because all possible resolvents are asymmetric tautologies:
\begin{eqnarray*}
F \land (a) \land (\bar b) \land (c) \upone \emptys &~~~\mathrm{using}~~~&(a \lor b \lor \bar c)\\
F \land (a) \land (\bar c) \land (\bar d) \upone \emptys &~~~\mathrm{using}~~~&(a \lor c \lor d)\\
F \land (a) \land (b) \land (d) \upone \emptys &~~~\mathrm{using}~~~&(a \lor \bar b \lor \bar d)
\end{eqnarray*}
\end{example}

\section{Cube-and-Conquer Solving}

Arguably the most effective method to solve many hard combinatorial problems via SAT technology is
the {\em cube-and-conquer} paradigm~\cite{HKWB11}, abbreviated by \CaC, due to strong performance and easy parallelization,
which has been demonstrated by the \CaC{} solver {\tt Treengeling}~\cite{biere2013} in recent SAT Competitions.
\CaC{} consists of two
phases. In the first phase, a look-ahead SAT solver~\cite{HvM09HBSAT} partitions the problem into many (potentially millions of)
subproblems. These subproblems, expressed as ``cubes'' (conjunctions) of the decisions (the literals set to true), are solved using a CDCL
 solver, also known as the ``conquer'' solver. The intuition behind this combination of paradigms is that look-ahead heuristics focus on global decisions, while CDCL heuristics
focus on local decisions.  Global decisions are important to split the problem,
while local decisions are effective when there exist a short refutation. So the
idea behind \CaC{} is to partition the problem until a short
refutation arises. \CaC{} can solve hard problems much faster than
either pure look-ahead or pure CDCL.  The problem with pure look-ahead solving is
that global decisions become poor decisions when a short refutation is present,
while pure CDCL tends to perform rather poor when there exist no short
refutation.  We will demonstrate that \CaC{} outperforms pure CDCL
and pure look-ahead in Section~\ref{sec:runtimes}. Apart from improved performance on a single core, \CaC{} allows for
easy parallelization.  The subproblems are solved independently, so they are
distributed on a large cluster. 

There are two \CaC{} variants: solving one cube per solver and
solving multiple cubes by an incremental solver. The first approach
allows solving cubes in parallel, while the second approach allows for reusing
heuristics and learned clauses while solving multiple cubes. The second approach
works as follows: an incremental SAT solver receives the input
formula and a sequence of cubes\footnote{In practice this is done using a single incremental CNF
file. For details about the format, see \url{http://www.siert.nl/icnf/}.}. After solving the formula
under the assumption that a cube is true, the solver does not terminate, but starts working on
a next cube. The heuristics and the learned clause database are not reset when starting solving a new cube, 
but reused to potentially exploit similarities between cubes.

In our computation we combined them, via a two-staged splitting, to exploit both parallelism and reusage. 
First the problem is split into $10^6$ cubes, and then for each cube, the corresponding subproblem is split
again creating billions of sub-cubes. An incremental SAT solver solves all the sub-cubes generated from a 
single cube sequentially. 


\section{Solving the boolean Pythagorean Triples Problem}

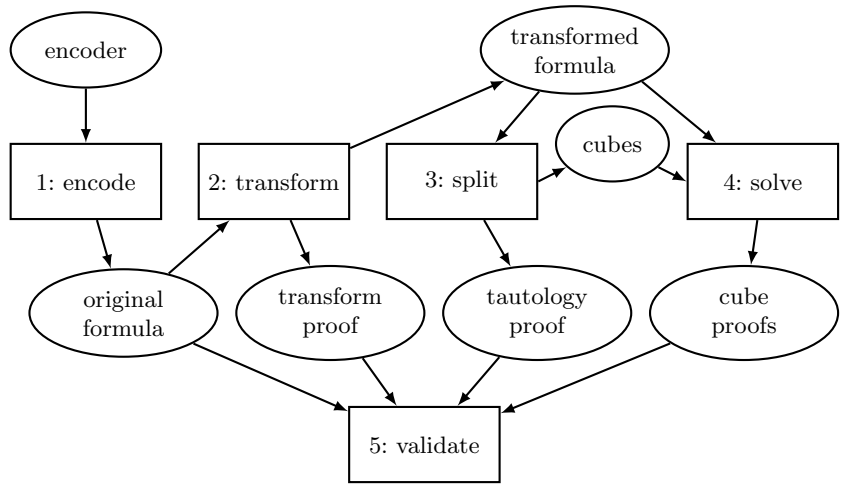
\begin{figure}[t]
\centering

\begin{tikzpicture}[every text node part/.style={align=center},thick,>=latex]
\node[draw,rectangle, minimum height=10mm, minimum width=20mm] (encode) at (0,-0.25) {1: encode};
\node[draw,rectangle, minimum height=10mm, minimum width=20mm] (transform) at (2.5,-0.25) {2: transform};
\node[draw,rectangle, minimum height=10mm, minimum width=20mm] (split) at (5,-0.25) {3: split};
\node[draw,rectangle, minimum height=10mm, minimum width=20mm] (solve) at (9,-0.25) {4: solve};
\node[draw,rectangle, minimum height=10mm, minimum width=20mm] (validate) at (4.5,-3.75) {5: validate};

\node[draw,ellipse, minimum height=10mm, minimum width=15mm] (cubes) at (7,0.25) {\!\!\!cubes\!\!\!};

\node[draw,ellipse, minimum height=10mm, minimum width=20mm] (encoder) at (0,1.5) {\!\!\!encoder\!\!\!};
\node[draw,ellipse, minimum height=10mm, minimum width=25mm] (original) at (0.5,-2) {\!\!\!original\!\!\!\\\!\!\!formula\!\!\!};
\node[draw,ellipse, minimum height=10mm, minimum width=25mm] (optimized) at (6.5,1.5) {\!\!\!transformed\!\!\!\\\!\!\!formula\!\!\!};

\node[draw,ellipse, minimum height=10mm, minimum width=25mm] (tproof) at (3.25,-2) {\!\!\!transform\!\!\!\\\!\!\!proof\!\!\!};

\node[draw,ellipse, minimum height=10mm, minimum width=25mm] (tautproof) at (6.0,-2) {\!\!\!tautology\!\!\!\\\!\!\!proof\!\!\!};

\node[draw,ellipse, minimum height=10mm, minimum width=25mm] (cproof) at (8.75,-2) {\!\!\!cube\!\!\!\\\!\!\!proofs\!\!\!};

\draw[->] (encoder) -- (encode);
\draw[->] (encode) -- (original);
\draw[->] (original) -- (transform);

\draw[->] (transform) -- (optimized);
\draw[->] (transform) -- (tproof);

\draw[->] (optimized) -- (split);
\draw[->] (optimized.south east) -- (solve);

\draw[->] (split) -- (tautproof);
\draw[->] (solve) -- (cproof);

\draw[->] (cproof) -- (validate);
\draw[->] (tproof) -- (validate);
\draw[->] (tautproof) -- (validate);
\draw[->] (original) -- (validate);

\draw[->] (split.east) -- (cubes);
\draw[->] (cubes) -- (solve.west);

\end{tikzpicture}
\vspace{-5pt}
\caption{Illustration of the framework to solve hard combinatorial problems. The phases 
are shown in the rectangle boxes, while the input and output files for these phases are 
shown in oval boxes.}
\label{fig:framework}
\end{figure}

Our framework for solving hard problems consists of five phases: encode, transform, split, solve, and validate. 
The focus of the encode phase is to make sure that representation of the problem as SAT instance is valid. 
The transform phase reformulates the problem to reduce the computation costs of the 
later phases. The split phase partitions the transformed formula into many, possibly millions of subproblems.
The subproblems are tackled in the solve phase. The validation phase checks whether the 
proofs emitted in the prior phases are a valid refutation for the original formula. Figure~\ref{fig:framework}
shows an illustration of the framework. The framework, including the specialized heuristics, have 
been developed by the first author, who also performed all implementations and experiments.

\subsection{Encode}

The first phase of the framework focusses on making sure that the problem
to be solved is correctly represented into SAT. In the second phase the representation 
will be optimized. The DRAT proof format can express all transformations. 

Formula $F_n$ expresses whether the natural numbers up to $n$ can be 
partitioned into two parts with no part containing a triple $(a,b,c)$ such that 
$a^2 + b^2 = c^2$. One set will be called the positive part, while the other 
will be called the negative part.
$F_n$ uses Boolean variables $x_i$ with $i \in \{1, \dots, n\}$. The assignment
$x_i$ to true / false, expresses that $i$ occurs in the positive / negative part, respectively. 
For each triple $(a,b,c)$ such that $a^2 + b^2 = c^2$, there is a constraint 
$\textsc{NotEqual} (a,b,c)$ in $F_n$, or in clausal form: 
$(x_a \lor x_b \lor x_c) \land (\bar x_a \lor \bar x_b \lor \bar x_c)$.

\subsection{Transform}

The goal of the transformation phase is to massage the initial encoding to 
execute the later phases more efficiently. A proof for the 
transformations is required to ensure that the changes are valid. Notice that 
a transformation that would be helpful for one later phase, might be harmful
for another phase. Selecting transformations is therefore
typically a balance between different trade-offs. For example, bounded 
variable elimination~\cite{SatELite} is a preprocessing technique that
tends to speed up the solving phase. However, this technique is generally
harmful for the splitting phase as it obscures the look-ahead heuristics. 

We applied two transformations.
First, blocked clause elimination (BCE)~\cite{BCE}. BCE on $F_{7824}$ and $F_{7825}$ has the following effect:
Remove $\textsc{NotEqual} (a,b,c)$ if $a$, $b$, or $c$ occurs only in this constraint,
and apply this removal until fixpoint. Note that removing a constraint $\textsc{NotEqual} (a,b,c)$
because e.g.\ $a$ occurs once, reduces the occurrences of $b$ and $c$ by one, and as a result $b$ or $c$ may occur only once after the removal, allowing further
elimination. We remark that a solution for the formula after the transformation may not satisfy
the original formula, however this can be easily repaired~\cite{BCE}. The numerical effects of this reduction are as follows: $F_{7824}$ has $6492$ (occurring) variables and $18930$ clauses, $F_{7825}$ has $6494$ variables and $18944$ clauses, while after BCE-reduction we get $3740$ variables and $14652$ clauses resp.\ $3745$ variables and $14672$ clauses.

The second transformation is symmetry breaking~\cite{Crawford}.
The Pythagorean Triples encoding has one symmetry: the two parts are 
interchangeable. To break this, we can pick an arbitrary variable $x_i$
and assign it to true (or, equivalently, put in the positive part). In practice it is best to pick the
variable $x_i$ that occurs most frequently in $F_n$. For the two formulas used during
our experiments, the most occurring variable is $x_{2520}$ which was used for symmetry breaking.
Symmetry breaking can be expressed in the DRAT format, but it is tricky. A
recent paper~\cite{sbp} explains how to construct this part of the transformation proof.

Bounded variable elimination (a useful transformation in
general) was not applied. Experiments showed that this transformation slightly increased the solving times. 
More importantly, applying bounded variable elimination transforms the problem into a non-3-SAT formula, 
thereby seriously harming the look-ahead heuristics, as the specialized 3-SAT heuristics can no longer be used.



\subsection{Split}

\begin{figure}[ht]
\centering
\begin{minipage}{0.45\textwidth}
\begin{tikzpicture}
\node[draw,circle] (root) at (0,3) {$\!x_5\!$};
\node[draw,circle] (n1) at (1.5,2.35) {$\!x_2\!$};
\node[draw,circle] (n2) at (-1.5,2.35) {$\!x_3\!$};
\node[draw,circle, fill=black] (n3) at (-1.95,1.5) {$~$};
\node[draw,circle] (n4) at (-0.75,1.4) {$\!x_7\!$};
\node[draw,circle, fill=black] (n5) at (-0.3,0.55) {$~$};
\node[draw,circle, fill=black] (n6) at (-1.2,0.55) {$~$};
\node[draw,circle, fill=black] (n7) at (1.05,1.5) {$~$};
\node[draw,circle] (n8) at (2.25,1.4) {$\!x_3\!$};
\node[draw,circle] (n9) at (1.5,0.45) {$\!x_6\!$};

\node[draw,circle, fill=black] (n10) at (2.7,0.55) {$~$};

\node[draw,circle, fill=black] (n11) at (1.05,-0.4) {$~$};
\node[draw,circle, fill=black] (n12) at (1.95,-0.4) {$~$};

\draw[-] (root) -- (n1) node [midway, above] {$\mathrm{\bf f}$};
\draw[-] (root) -- (n2) node [midway, above] {$\mathrm{\bf t}$};

\draw[-] (n2) -- (n3) node [midway, left] {$\mathrm{\bf f}$};;
\draw[-] (n2) -- (n4) node [midway, right] {$\mathrm{\bf t}$};;

\draw[-] (n4) -- (n5) node [midway, right] {$\mathrm{\bf f}$};
\draw[-] (n4) -- (n6) node [midway, left] {$\mathrm{\bf t}$};

\draw[-] (n1) -- (n7) node [midway, left] {$\mathrm{\bf t}$};

\draw[-] (n1) -- (n8) node [midway, right] {$\mathrm{\bf f}$};
\draw[-] (n8) -- (n9) node [midway, left] {$\mathrm{\bf t}$};

\draw[-] (n8) -- (n10) node [midway, right] {$\mathrm{\bf f}$};

\draw[-] (n9) -- (n11) node [midway, left] {$\mathrm{\bf f}$};
\draw[-] (n9) -- (n12) node [midway, right] {$\mathrm{\bf t}$};

\end{tikzpicture}
\end{minipage}
~~~~~~~~
\begin{minipage}{0.35\textwidth}
{~~cube file in {\tt inccnf} format}
\begin{Verbatim}[frame=single]
  a  5 -3  0
  a  5  3  7  0
  a  5  3 -7  0
  a -5  2  0
  a -5 -2  3 -6  0
  a -5 -2  3  6  0
  a -5 -2 -3  0
\end{Verbatim}
\end{minipage}
\caption{A binary branching tree (left) with the decision variables in the nodes and the polarity on the edges. The corresponding cube file (right) in the {\tt inccnf} format. The prefix {\tt a} denotes {\em assumptions}. Positive numbers express positive literals, while negative numbers
express negative literals. Each cube (line) is terminated with a {\tt 0}.}
\label{fig:tree}
\end{figure}
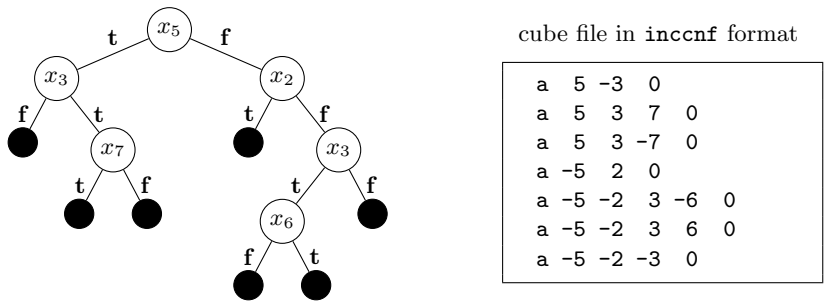

Partitioning is crucial to solve hard combinatorial problems. Effective partitioning is based
on global heuristics~\cite{HKWB11} --- in contrast to the ``local'' heuristics used in CDCL solvers. 
The result of partitioning is a binary branching tree of which the leaf nodes represent a subproblem
of the original problem. The subproblem is constructed by adding the conjunction of decisions
that lead to the leaf as unit clauses. Figure~\ref{fig:tree} shows such a partitioning as a 
binary tree with seven leaf nodes (left) and the corresponding list of seven cubes (right). The cubes
are shown in the {\tt inccnf} format that is used for incremental solvers to guide their 
search. 

Splitting heuristics are crucial in solving a problem efficiently.
In practice, the best heuristics are based on {\em look-aheads}~\cite{HvM09HBSAT,Kullmann2007HandbuchTau}. In short, a look-ahead refers 
to assigning a variable to a truth value followed by unit propagation and measuring the 
changes to the formula during the propagation. It remains to find good measures.
The simplest measure is to count the number of assigned variables; measures
like that can be used for tie-breaking, but as has been realised in the field of
heuristics~\cite{Kullmann2007HandbuchTau}, the expected future gains for
unit-clause propagation, given by \emph{new short clauses}, are more important
than the current reductions.
The default heuristic in \CaC, which works well on most
hard-combinatorial problems, weighs all new clauses using
weights based on the length of the new clause (with an exponential decay of
the weight in the length). However for our Pythagorean Triples encoding,
using a refinement coming from random 3-SAT turned out to be more powerful.
Here all newly created clauses are binary, i.e., ternary clauses that
become binary during the look-ahead. The weight of a new binary clause depends on the occurrences of its two literals in the formula, estimating how likely
they become falsified.
This better performance is not very surprising as the formulas $F_n$ exhibit somewhat akin behavior 
to random 3-SAT formulas: i) all clauses have length three; and ii) the distribution of the 
occurrences of literals is similar. On the other hand, $F_n$ consists of clauses
with either only positive literal or only negative literals --- in contrast to random 3-SAT.

\subsection{Details regarding the heuristics}
\label{sec:heuristics}

The heuristics used for splitting extends the \emph{recursive weight heuristics}~\cite{MijndersWildeHeule2010March}, based on earlier work \cite{LA1996,Li1999Satz,DuboisDequen2001Kcnfs,DequenDubois2003Kcnfs}, by introducing minimal and maximal values $\alpha, \beta$, and choosing different parameters, optimized for the special case at hand.
A look-ahead on literal $l$ measures the difference between a formula before and after 
assigning $l$ to true followed by simplification. Let $F$ (or $F_l$) denote the formula 
before (or after) the look-ahead on $l$, respectively. We assume that $F$ and $F_l$ are
fully simplified using unit propagation. Thus $F_l \setminus F$ is the set of new
clauses, and the task is to weigh them; we note that each clause in $F_l \setminus F$ is binary. Each literal is assigned a heuristic value $h(l)$ and the weight $w_{y \lor z}$
for $(y \lor z) \in F_l \setminus F$ is defined as $h(\bar y) \cdot h(\bar z)$. The values of $h(l)$ are
computed using multiple iterations $h_0(l), h_1(l), \dots$, choosing the level with optimal performance, balancing the predictive power of the heuristics versus the cost to compute it. The idea of the heuristic values $h_i(l)$ is to approximate how strongly the literal $l$ is forced to true by the clauses containing $l$ (via unit propagation).
First, for all literals $l$, $h_0(l)$ is initialized to 1: $h_0(x) = h_0(\bar x) = 1$.
At each level $i \ge 0$, the average value $\mu_i$ is computed in order to scale the heuristics values $h_i(x)$:
\begin{equation}
\mu_i = \frac{1}{2n} \sum_{x \in \var({F})} \big( h_i(x) + h_i(\bar x) \big).
\end{equation}

Finally, in each next iteration, the heuristic values $h_{i+1}(x)$ are computed in which literals 
$y$ get weight ${h_{i}(\bar y)}/{\mu_i}$. The weight $\gamma$ expresses the relative importance of binary
clauses. This weight could also be seen as the heuristic value of a falsified literal.
Additionally, we have two other parameters, $\alpha$ expressing the minimal heuristic value and
$\beta$ expressing maximum heuristic value.
\begin{equation}
h_{i+1}(x) = \max(\alpha, \min(\beta, \sum_{(x \lor y \lor z) \in {F}} \Big(\frac{h_i(\bar y)}{\mu_i} \cdot \frac{h_i(\bar z)}{\mu_i}\Big) + 
			\gamma \!\!\!\! \sum_{(x \lor y) \in {F}} \frac{h_i(\bar y)}{\mu_i} )).
\end{equation}

In each node of the branching tree we compute $h(l) := h_4(l)$ for all literals occurring in the formula.
We use $\alpha = 8$, $\beta = 550$, and $\gamma = 25$. The ``magic" constants
differ significantly compared to the values used for random 3-SAT formulas where 
$\alpha = 0.1$, $\beta = 25$, and $\gamma = 3.3$ appear optimal~\cite{MijndersWildeHeule2010March}. The branching variable $x$ chosen is a variable with maximal $H(x) \cdot H(\bar x)$, where $H(l) := \sum_{y \lor z \in F_l \setminus F} w_{y \lor z}$.



\subsection{Solve}

The solving phase is the most straightforward part of the framework. It takes the transformed formula
and cube files as input and produces a proof of unsatisfiability of the transformed formula. Two different
approaches can be distinguished in general: one for ``easy'' problems and one for ``hard" problems. A problem is 
considered easy when it can be solved in reasonable time, say within a day on a single core. In that 
case, a single cube file can be used and the incremental SAT solver will emit a single proof file. The
more interesting case is when problems are hard and two levels of splitting are required.

The boolean Pythagorean triples problem $F_{7825}$ is very hard and required two level splitting:
the total runtime was approximately $4$ CPU years (21,900 CPU hours for splitting and 13,200 CPU hours for solving). 
Any problem requiring that amount of resources has to be solved in parallel. The first level 
consists of partitioning the problem into $10^6$ subproblems, which required approximately 1000 seconds on a single core;
for details see Section~\ref{sec:runtimes}. Each subproblem is represented by a cube $\varphi_i$ with $i \in \{1, \dots, 10^6 \}$ expressing a
conjunction of decisions. On the second level of splitting, each subproblem $F_{7825} \land \varphi_i$
is partitioned again using the same look-ahead heuristics. In contrast to the first level, the cubes generated 
on the second level are not used to create multiple subproblems. Instead, the second level cubes are 
provided to an incremental SAT solver together with a subproblem $F_{7825}$ and assumptions $\varphi_i$. 
The second level cubes are used to guide the CDCL solver. The advantage of guiding the CDCL solver is that learned
clauses computed while solving one cube can be reused when solving another cube. 

For each subproblem $F_{7825} \land \varphi_i$, the SAT solver produces a DRAT refutation. Most state-of-the-art
SAT solvers currently support the emission of such proofs. One can check that the emitted proof of
unsatisfiability is valid for $F_{7825} \land \varphi_i$. In this case, no changes to the proof logging of the solver
are required. However, in order to create an unsatisfiability proof of $F_{7825}$ by concatenating the proofs of
subproblems, all lemmas generated while solving $F_{7825} \land \varphi_i$ need to be extended with 
the clause $\neg \varphi_i$, and the SAT solver must not delete clauses from $F_{7825}$.


\subsection{Validate}

The last phase of the framework validates the results of the earlier phases. 
First, the encoding into SAT needs to be validated. This can be done by proving that the 
encoding tool is correct using a theorem prover. Alternatively,
a small program can be implemented whose correctness can be checked manually.
For example, our encoding tool consists of only 19 lines of C code. 
For details and validation files, check out \url{http://www.cs.utexas.edu/~marijn/ptn/}.

The second part consists of checking the three types of DRAT proofs produced in the earlier 
phases: the transformation, tautology, and the cube proofs. DRAT proofs can 
be merged easily by concatenating them. The required order for merging the 
proofs is: transformation proof, cube proofs, and tautology proof. 


\vspace{-10pt}

\subsubsection*{Transformation Proof} The transformation proof expresses how the 
initial formula, created by the encoder, is converted into a formula that is easier to solve. 
This part of the proof is typically small. The latest version of the {\tt drat-trim} checker
supports validating transformation proofs without requiring the other parts of the proof, 
based on a compositional argument~\cite{comproof}.

\vspace{-10pt}

\subsubsection*{Cube Proofs}

The core of the validation is checking whether the negation of each cube, the clause $\neg \varphi_i$, is 
implied by the transformed formula. Since we partitioned the problem using $10^6$ cubes, 
there are $10^6$ of cube proofs. We generated and validated them all. However, their 
total size is too large to share: almost 200 terabyte in the DRAT format. We tried to compress the
proof using a range of dedicated clause compression techniques~\cite{ziproof} combined with 
state-of-the-art general purpose tools, such as {\tt bzip2} or {\tt 7z}. After compression
the total proof size was still 14 terabytes. So instead we provide the cube files for the subproblems as a certificate. Cube files can
be compressed heavily, because they form a tree. Instead of storing all cubes as a list of
literals, shown as in Figure~\ref{fig:tree}, it is possible to store only one literal per cube. 
Storing the literal in a binary format~\cite{ziproof} followed by {\tt bzip2} allowed us to 
store all the cube files using ``only" 68 gigabytes of disk space. 
We added support for the {\tt inccnf} format to {\tt glucose} 3.0 in order to solve the cube files.
This solver can also reproduce the DRAT proofs in about $13,\!000$ CPU hours. 
Checking these proofs requires about $16,\!000$ CPU hours, so reproducing the DRAT proofs 
almost doubles the validation effort. This is probably
a smaller burden than downloading and decompressing many terabytes of data. 

%
%

\vspace{-10pt}

\subsubsection*{Tautology Proof}

A cube partitioning is valid, i.e., covers the complete search space, if the disjunction of cubes
is a tautology. This needs to be checked during the validation phase. Checking this
can be done by negating the disjunction of cubes and feed the result to a CDCL solver which
supports proof logging. If the solver can refute the formula, then the disjunction of cubes is a tautology.
We refer to the proof emitted by the CDCL solver as the {\em tautology proof}. This tautology proof
is part of the final validation effort.

%
%
%
%
%
%
%
%
%
%
%


\section{Results}

This section offers details of solving the boolean Pythagorean Triples problem\footnote{Files and 
tools can be downloaded at \url{http://www.cs.utexas.edu/~marijn/ptn/}}. 
All experiments were executed on the Stampede cluster\footnote{\url{https://www.tacc.utexas.edu/systems/stampede}}.
Each node on this cluster consists of an Intel Xeon Phi 16-core CPU and 32 Gb memory. 
We used cube solver {\tt march\_cc} and conquer solver {\tt glucose} 3.0 during our experiments.

\vspace{-10pt}

\subsection{Heuristics}

In our first attempt to solve the Pythagorean triples problem, we partitioned the problem (top-level and subproblems) using the default decision heuristic in the 
cube solver {\tt march\_cc} for 3-SAT formulas. After some initial experiments, we estimated that the total runtime of solving (including splitting) $F_{7825}$ would be roughly $300,\!000$ CPU 
hours on the Stampede cluster. To reduce the computation costs, we (manually) optimized the magic constants in {\tt march\_cc}, resulting in the
heuristic presented in Section~\ref{sec:heuristics}. The new heuristics reduced the total runtime to $35,\!000$ CPU hours, so by almost an order of magnitude.  
Table~\ref{tab:heuristics} shows the results of various heuristics on five randomly selected subproblems. Here, we
optimized {\tt march\_cc} in favor of the other heuristics to make the comparison more fair: we turned off look-ahead preselection, which
is helpful for the new heuristics (and thus used in the computation), but harmful for the other heuristics.

\begin{table}[t]
\caption{Solving times for \CaC{} using different look-ahead heuristics and pure CDCL. The top left, bottom left, and right numbers expresses the cube, conquer, and their sum times,
respectively. {\em Ptn 3-SAT} is 3-SAT heuristics optimized for Pythagorean triple problems; {\em rnd 3-SAT} is the 3-SAT heuristics optimized for random 3-SAT (default); {\em \#bin} is the sum of new binary clauses; and  {\em \#var} is the number of assigned variables.}
\label{tab:heuristics}
\centering
\begin{tabular}{l|rr|rr|rr|rr|r}
{\em cube \#} & \multicolumn{2}{c|}{\em Ptn 3-SAT} & \multicolumn{2}{c|}{\em rnd 3-SAT} & \multicolumn{2}{c|}{\em \#bin} & \multicolumn{2}{c|}{\em \#var} & \multicolumn{1}{c}{\em pure CDCL}\\ \hline
\multirow{ 2}{*}{104302} & ~152.98~ & \multirow{ 2}{*}{\bf 228.48\,} & ~608.46~ & \multirow{ 2}{*}{783.40\,} & ~263.23~ & \multirow{ 2}{*}{413.94\,} & ~789.43~ & \multirow{ 2}{*}{1053.22\,} &  \multirow{ 2}{*}{1372.87}~\\ 
                                               &  ~75.50~  & &  ~174.94~  & &  ~150.71~  & &  ~263.79~ &   \\ \hline
\multirow{ 2}{*}{268551} & ~74.03~ & \multirow{ 2}{*}{\bf 107.86\,} & ~92.09~ & \multirow{ 2}{*}{140.91\,} & ~98.93~ & \multirow{ 2}{*}{154.76\,} & ~487.45~ & \multirow{ 2}{*}{707.72\,} &  \multirow{ 2}{*}{150.06}~\\ 
                                               &  ~33.83~  & &  ~48.82~  & &  ~55.83~  & &  ~220.27~  & \\ \hline
\multirow{ 2}{*}{934589} & ~136.94~ & \multirow{ 2}{*}{\bf 211.38\,} & ~206.28~ & \multirow{ 2}{*}{328.27\,} & ~156.78~ & \multirow{ 2}{*}{263.94\,} & ~529.21~ & \multirow{ 2}{*}{764.91\,} &  \multirow{ 2}{*}{631.91}~\\ 
                                               &  ~74.44~  & &  ~121.99~  & &  ~107.16~  & &  ~235.70~  & \\ \hline                                               
\multirow{ 2}{*}{950025} & ~143.69~ & \multirow{ 2}{*}{\bf 217.78\,} & ~152.49~ & \multirow{ 2}{*}{252.16\,} & ~203.18~ & \multirow{ 2}{*}{341.27\,} & ~550.47~ & \multirow{ 2}{*}{777.46\,} &  \multirow{ 2}{*}{330.61}~\\ 
                                               &  ~74.09~  & &  ~99.67~  & &  ~138.09~  & &  ~226.99~  & \\ \hline
\multirow{ 2}{*}{980757} & ~112.22~ & \multirow{ 2}{*}{\bf 142.63\,} & ~170.34~ & \multirow{ 2}{*}{224.24\,} & ~181.14~ & \multirow{ 2}{*}{241.67\,} & ~685.04~ & \multirow{ 2}{*}{845.97\,} &  \multirow{ 2}{*}{155.57}~ \\ 
                                               &  ~30.41~  & &  ~53.90~  & &  ~60.53~  & &  ~160.93~ &  \\ \hline
\end{tabular}
\end{table}

\vspace{-10pt}

\subsection{Cube and Conquer}
\label{sec:runtimes}

The first step of the solving phase was partitioning the transformed formula into many subproblems using
look-ahead heuristics. Our cluster account allowed for running on $800$ cores in parallel. 
We decided to partition the problem into a multiple of $800$ to perform easy parallel execution: exactly $10^6$. 
Partitioning the formula into $10^6$ subproblems ensured that the conquer time of solving
most subproblems is less than two minutes, a runtime with the property that proof validating can be achieved 
in a time similar to the solving time. 

\begin{figure}[h!]
~~~\,\includegraphics[width=.45\textwidth]{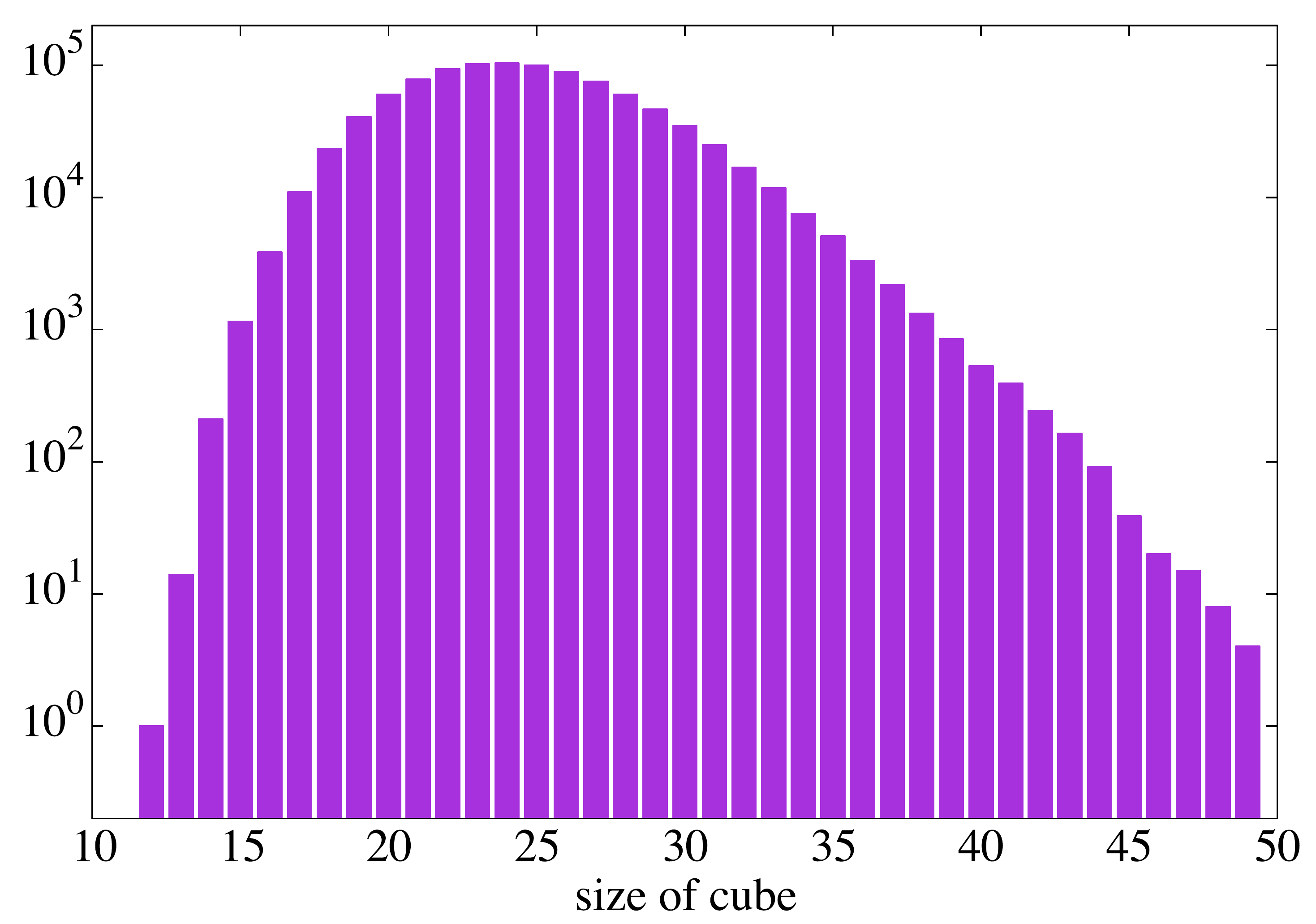}
\hfill
\includegraphics[width=.45\textwidth]{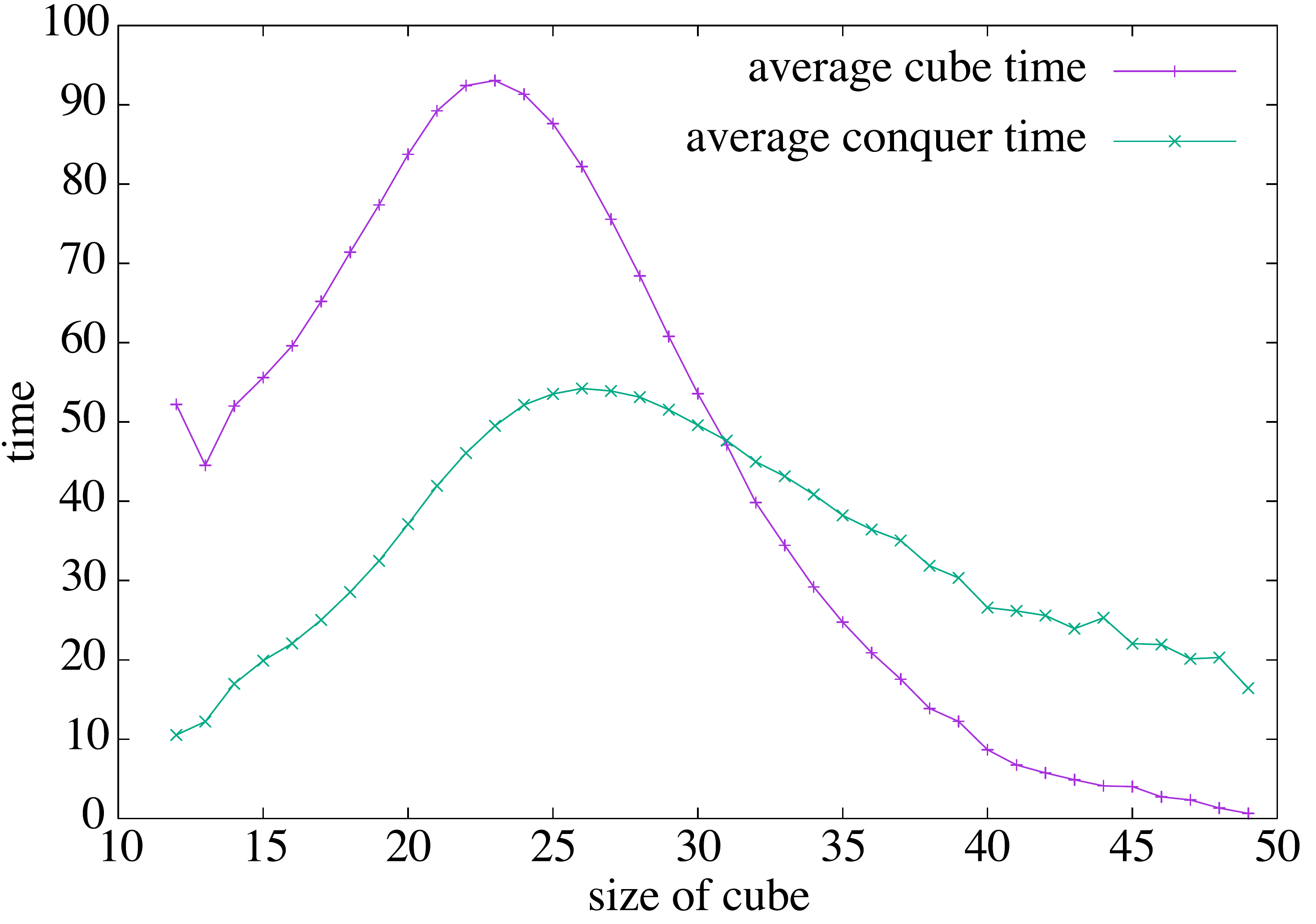}
\vspace{-10pt}
\caption{Left, a histogram (logarithmic) of the cube size of the $10^6$ subproblems. 
Right, average runtimes per size for the split (cube) and solve (conquer) phases.}
\label{fig:hist}
\end{figure}

A simple way of splitting a problem into $10^6$ subproblems is to build a balanced binary branching tree of depth 20. 
However, using a balanced binary branching tree results in 
poor performance on hard combinatorial problems~\cite{HKWB11}. A more effective partitioning heuristic picks the leaf
nodes such that the number of assigned variables (including inferred variables) in those nodes are equal. Based on some initial experiments, we observed that the best heuristics  for Pythagorean Triples formulas however is to count the number of binary
clauses in each node. Recall that all clauses in the transformed formula are ternary. Selecting nodes in 
the decision tree that have about $3,000$ binary clauses resulted in $10^6$ subproblems. Figure~\ref{fig:hist} 
(left) shows a histogram of the depth of the branching tree (or, equivalently, the size of the cube) of the selected nodes.
Notice that the smallest cube has size 12 and the largest cubes have size 49. 

Figure~\ref{fig:hist} (right) shows the time for the cube and conquer runtimes averaged per size of the cubes. 
The peak average of the cube runtime is around size $24$, while the peak of the conquer runtime is around size $26$. 
The cutoff heuristics of the cube solver for second level splitting were based on the number of unassigned variables, 
$3450$ variables to be precise.

\begin{figure}[h!]
~\includegraphics[width=.46\textwidth]{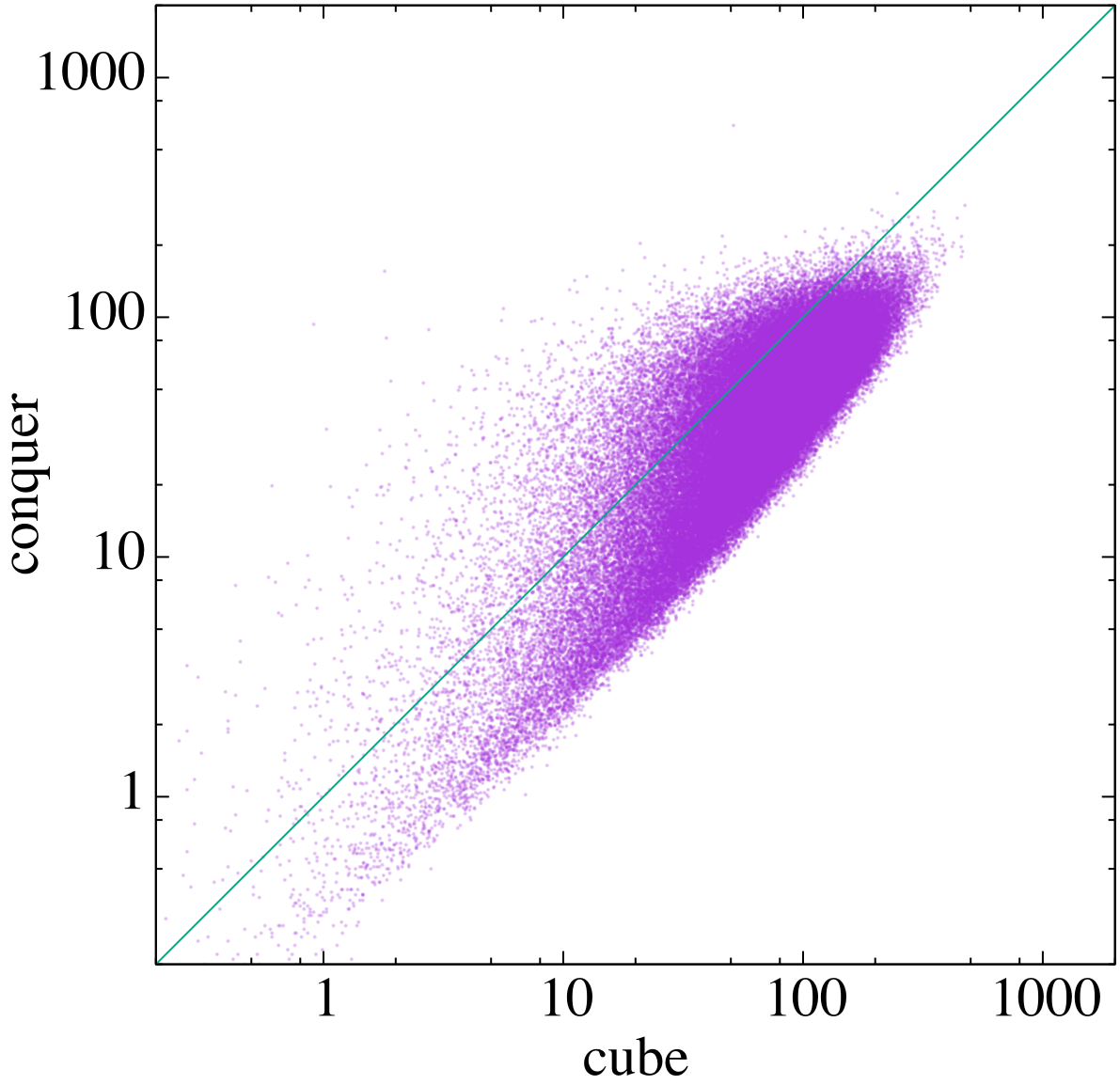}
\hfill
\includegraphics[width=.46\textwidth]{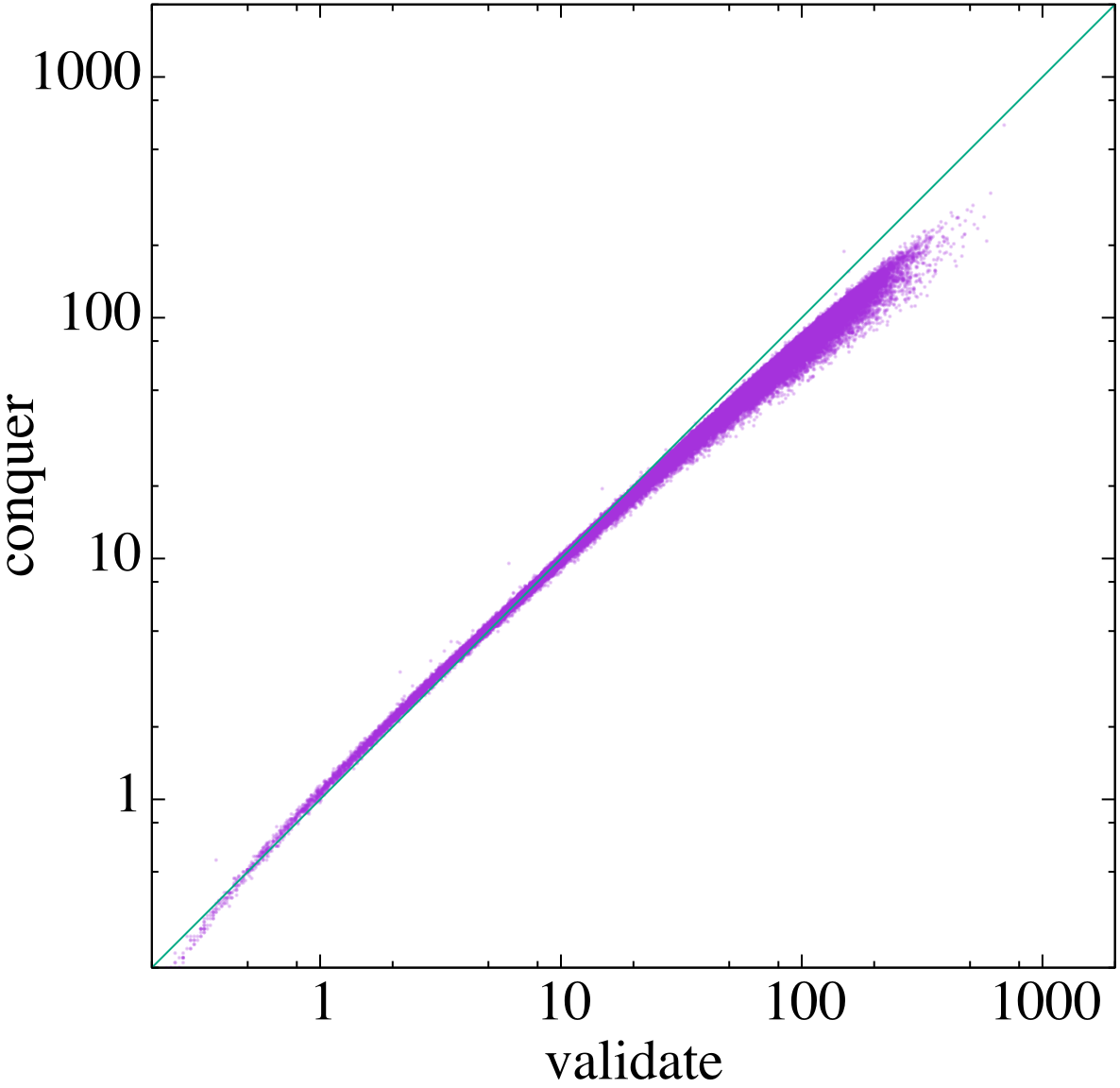}~~
\vspace{-8pt}
\caption{Left, a scatter plot comparing the cube (split) and conquer (solve) time per subproblem. 
Right, a scatter plot comparing the validation and conquer time.}
\label{fig:scatter}
\end{figure}

A comparison between the cube, conquer, and validation runtimes is shown in Figure~\ref{fig:scatter}.
The left scatter plot compares cube and conquer runtimes. It shows that within our experimental setup 
 the cube computation is about twice as expensive compared to the conquer computation. The right
scatter plot compares the validation and conquer runtimes. It shows that these times are very similar. 
Validation runtimes grow slightly faster compared to conquer runtimes. The average cube, conquer, 
and validation times for the $10^6$ subproblems are $78.87$, $47.52$, and $60.62$ seconds, respectively.

\begin{figure}[h!]
~\includegraphics[width=.465\textwidth]{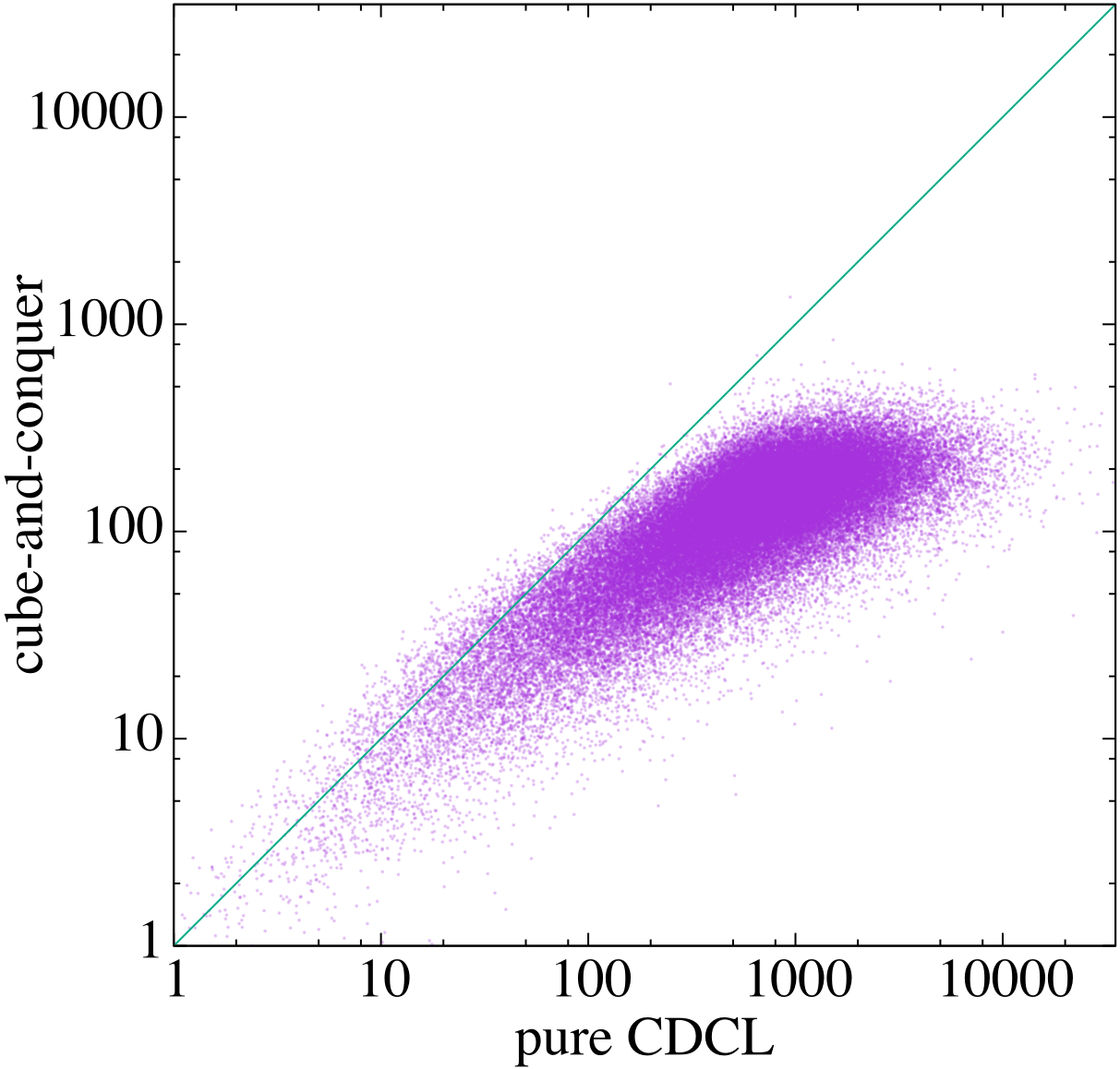}
\hfill
\includegraphics[width=.46\textwidth]{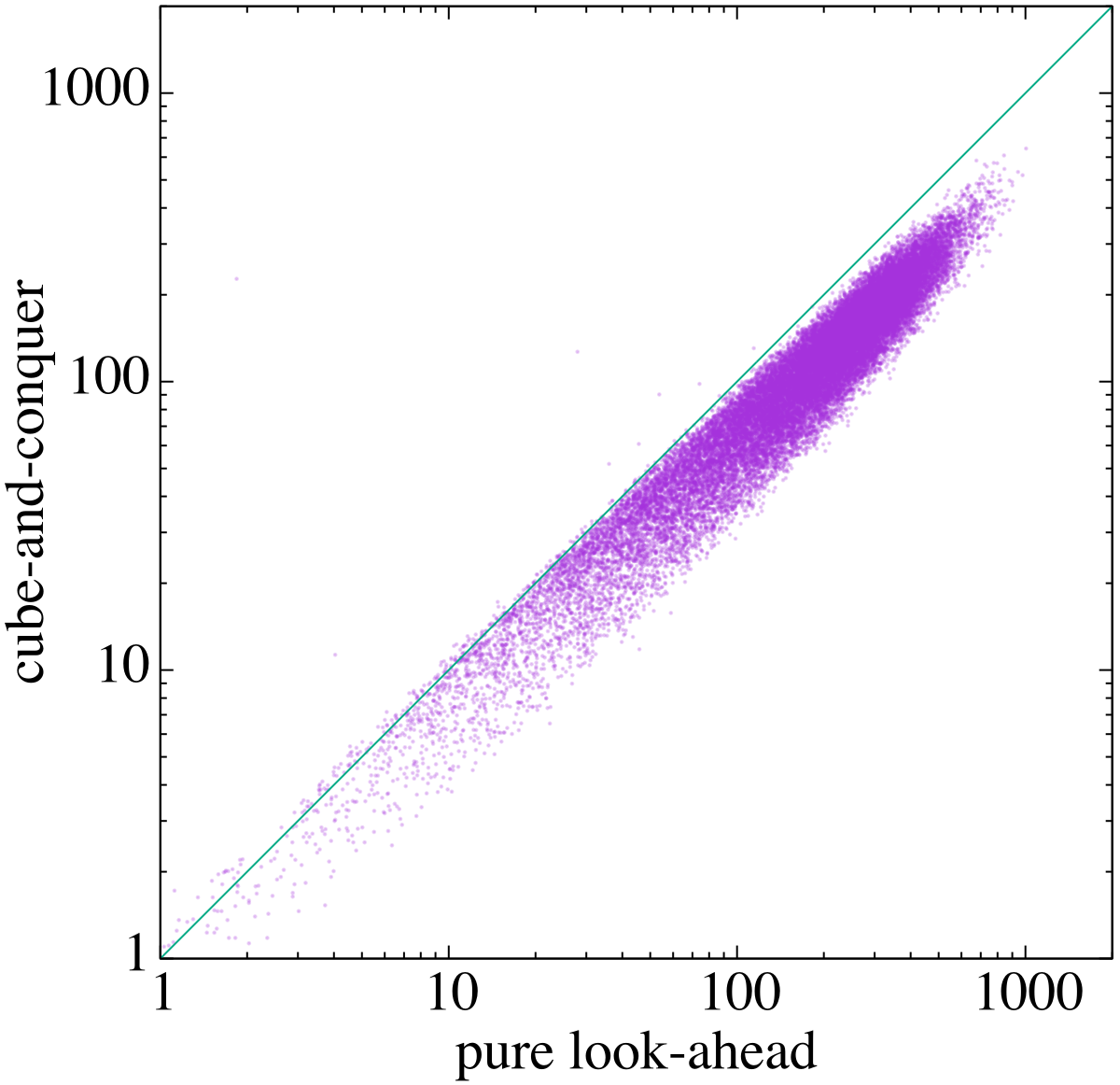}~~
\vspace{-8pt}
\caption{Scatterplots comparing cube-and-conquer to pure CDCL (left) and pure 
look-ahead (right) solving methods on the Pythagorean Triples subproblems.}
\label{fig:scatter-cnc}
\end{figure}

Figure~\ref{fig:scatter-cnc} compares the cube+conquer runtimes to solve the
$10^6$ subproblems with the runtimes of pure CDCL (using {\tt glucose} 3.0) and pure look-ahead
(using {\tt march\_cc}).
The plot shows that cube+conquer clearly outperforms pure CDCL.
Notice that no heuristics of {\tt glucose} 3.0 were changed during all experiments for
both cube+conquer and pure CDCL. In particular, a variable decay of 0.8 was used
throughout all experiments as this is the {\tt glucose} default. However, we observed
that a higher variable decay (in between 0.95 and 0.99) would improve
the performance of both cube+conquer and pure CDCL. We did not optimize 
 {\tt glucose} to keep it simple, and because the conquer part is already the
 cheapest phase of the framework (compared to split and validate); indeed
frequently speed-ups of two orders or magnitude could be achieved on the
harder instances. Pure look-ahead is also slower compared to cube+conquer,
but the differences are smaller: on average cube+conquer is about twice as 
fast.\vspace{-2ex}

%
%
%


\subsection{Extreme Solutions}
Of the $10^6$ subproblems that were created during the splitting phase,
only one subproblem is satisfiable for the extreme case, i.e., $n=7824$. This
suggests that the formula after symmetry breaking has a big {\em backbone}. A
variable belongs to backbone of a formula if it is assigned to the same
truth value in all solutions. We computed the backbone of $F_{7824}$, which
consists of $2304$ variables. 
The backbone reveals why it is impossible
to avoid Pythagorean Triples indefinitely when partitioning the natural numbers into two parts: variables $x_{5180}$ and $x_{5865}$ are both positive in the backbone, 
forcing $x_{7825}$ to be negative due to $5180^2 + 5865^2 = 7825^2$. At
the same time, variables $x_{625}$ and $x_{7800}$ are both negative in the
backbone forcing $x_{7825}$ to be positive due to $625^2 +
7800^2 = 7825^2$.

A satisfying assignment does not necessarily assign all natural numbers up to $7824$ 
that occur in Pythagorean Triples. For example, we found a satisfying assignment
that assigns only $4925$ out of the $6492$ variables occurring in $F_{7824}$.
So not only is $F_{7824}$ satisfiable, but it has a huge number of solutions.
\vspace{-1.5ex}

\section{Conclusions}

We solved and verified the boolean Pythagorean Triples problem using
\CaC. The total solving time was about $35,\!000$ hours
and the verification time about $16,\!000$ hours. Since \CaC{}
allows for massive parallelization, resulting in almost linear-time speedups,
the problem was solved altogether in about two days on the Stampede cluster. Apart from
strong computational resources, dedicated look-ahead heuristics were 
required to achieve these results. In future research we want to further 
develop effective look-ahead heuristics that will work for such hard 
combinatorial problems out of the box. We expect that parallel 
computing combined with look-ahead splitting heuristics will make it 
feasible to solve many other hard combinatorial problems that are too
hard for existing techniques. Moreover, we argue that solutions to 
such problems require certificates that can be validated by the community
--- similar to the certificate we provided for the  boolean Pythagorean Triples problem.
A fundamental question is whether Theorem \ref{thm:bPyth} has a ``mathematical''
(human-readable) proof, or whether the gigantic (sophisticated)
case-distinction, which is at the heart of our proof, is the best
there is?
It is conceivable that Conjecture \ref{con:Pyth} is true,
but for each $k$ has only proofs like our proof, where the size of
these proofs is growing so quickly, that Conjecture \ref{con:Pyth} is
actually not provable in current systems of foundations of Mathematics (like ZFC).

\paragraph*{\bf Acknowledgements}\!
The authors acknowledge the Texas Advanced Computing Center (TACC) at The
University of Texas at Austin for providing grid resources that have
\mbox{contributed to the research results reported within this paper.}

\bibliography{pyth,comproof}
\bibliographystyle{plain}

%
%
%


\newcommand{\bout}[2]{\fill[yellow] (#1,#2)  rectangle (#1+1,#2+1);}

\newcommand{\pin}[2]{\fill[purple] (#1,#2)  rectangle (#1+1,#2+1);}

\newcommand{\pout}[2]{\fill[xorange] (#1,#2)  rectangle (#1+1,#2+1);}

\newcommand{\nin}[2]{\fill[xblue] (#1,#2)  rectangle (#1+1,#2+1);}

\newcommand{\nout}[2]{\fill[xgreen] (#1,#2)  rectangle (#1+1,#2+1);}

\end{document}